\documentclass[]{aastex701}

\usepackage{accents}
\usepackage{makecell}
\usepackage{tabularx}
\usepackage{booktabs} 
\usepackage{graphicx}
\usepackage{caption}
\usepackage{hyperref}
\usepackage{soul}

\begin{document}

\title{The Birth of Be Star Disks
II. A High-Resolution Spectroscopic Campaign and TESS Observations of an Outburst of the Classical Be star $\lambda$ Pavonis }

\author[0009-0000-5406-8658]{Sola S. Nova}
\affiliation{Department of Physics and Astronomy, Embry-Riddle Aeronautical University, 
3700 Willow Creek Rd, 
Prescott, AZ 86301, USA}
\email{MILLERX@my.erau.edu}

\author[0000-0002-2806-9339]{Noel D. Richardson}
\affiliation{Department of Physics and Astronomy, Embry-Riddle Aeronautical University, 
3700 Willow Creek Rd, 
Prescott, AZ 86301, USA}
\email{noel.richardson@erau.edu}

\author[0000-0002-2919-6786]{Jonathan Labadie-Bartz}
\affiliation{DTU Space, Technical University of Denmark, Elektrovej 327, Kgs., Lyngby 2800, Denmark}
\affiliation{LIRA, Observatoire de Paris, Université PSL, CNRS, Sorbonne Université, Université Paris Cité, CY Cergy \\ Paris Université, 92190
Meudon, France}
\email{jbartz@udel.edu}

\author[0009-0008-0631-6926]{Samantha Garcia Flores}
\affiliation{Department of Physics and Astronomy, Embry-Riddle Aeronautical University, 
3700 Willow Creek Rd, 
Prescott, AZ 86301, USA}
\email{GARCIAFS@my.erau.edu}

\begin{abstract}
Be stars are rapidly-rotating B stars that have shown emission lines originating in a circumstellar disk. The mechanisms that lead to disk formation and dissipation are not known although progress has been made with some systems. We present a study of a disk outburst of the Be star $\lambda$ Pavonis. Our dataset comprises 698 high‑resolution spectra contemporaneous with TESS photometry in 2023. Near the end of TESS monitoring, the star began disk building from a pristine diskless state. We find the disk built within 5 days in optical \ion{H}{1} and \ion{He}{1} lines, while the disk circularized in about 12 days. The disk began to decay in higher excitation \ion{He}{1} first, then lower excitation transitions, with the decay ending last for H$\alpha$. We examine non-radial pulsations both through TESS photometry and the line profile variations (LPVs) in the spectroscopy. Our analysis indicates that two periodicities seen in TESS photometry (at 1.644 and 1.485 cycles d$^{-1}$) are not seen in the spectral lines before, during, or after the outburst. The strongest photometric signal is a periodicity at 0.163 cycles d$^{-1}$, which appears as a difference between the two weaker signals and is visible in the spectra without any apparent changes in amplitude or phase. We additionally find evidence for fast non-photometric pulsational variations over the course of spectroscopy obtained before, during, and after the outburst. These fast LPVs are strong, and interfere with the two weaker signals, hampering our ability to detect them in spectroscopy.
\end{abstract}


\section{Introduction}

{Be stars are rapidly rotating, non-supergiant B-type stars that have shown Balmer-line emission arising from a circumstellar Keplerian decretion disk \citep[e.g.,][]{Collins_1987,2013A&ARv..21...69R}. The disks are variable on short, observable time scales, which are well described by the Viscous Decretion Disk (VDD) framework \citep{2006ApJ...639.1081C,2008ApJ...684.1374C,1991MNRAS.250..432L}. One active area of observational research for understanding these stars is correlating various observable signatures such as line equivalent widths (EWs) and the continuum flux with the growth and dissipation of these disks. Recently, \citet{2025arXiv250407571L} (hereafter, Paper I) presented work for 30 disk outbursts in 13 stars that were observed both with spectroscopy and photometrically with the Transiting Exoplanet Survey Satellite \citep[TESS;][]{2015JATIS...1a4003R}. We present in this paper a more in-depth analysis of the Be star $\lambda$ Pavonis (HR 7074, HD 173948) that was reported on in Paper I. Our analysis extends to include five sectors of TESS photometry (sectors 13, 66, 67, 93, and 94) along with the $\sim700$ high-resolution spectra discussed in Paper I, with the generalized data reduction being described previously in Paper I. 
Paper I focused primarily on the H$\alpha$ behavior and the circularization\footnote{The term circularization may be ambiguous in general. In this work, we use this term to refer to the circumstellar material becoming azimuthally homogenized -- i.e. forming an axi-symmetric disk. We are not attempting to describe the early-time orbital paths of individual circumstellar particles, which may have elliptical or circular orbits.} process and documented the full spectroscopic dataset. 
In this paper, we focus on complementary analyses so that in addition to H$\alpha$, we study H$\beta$, \ion{He}{1} $\lambda$5876, $\lambda$6678, $\lambda$4921, $\lambda$4713 and \ion{Si}{3} $\lambda$5739, with these lines chosen as they are available with both spectrographs used, CHIRON on the {Cerro Tololo Inter-American Observatory} (CTIO) 1.5-m \citep{2013PASP..125.1336T} and NRES on the {Las Cumbres Observatory} (LCO) 1.0-m \citep{2014SPIE.9147E..16E}. Our expanded analysis includes (i) time-resolved EW and V/R (ratio of EW of violet side to EW of red side of the line) fitting for the seven aforementioned optical spectral lines, (ii) a pulsational study connecting multi-epoch TESS photometry with spectroscopic line-profile variability (LPV), and (iii) a high-cadence intra-night LPV characterization using nine dense sequences. The  datasets used are presented in Table \ref{tab:obs_summary}.}

Prior to the presentation of these data and our analysis, we {review} a few relevant observational and theoretical considerations related to Be stars and their disks. Several physical processes have been suggested to {elucidate} how the disks around Be stars are formed. For example, stellar winds confined to the equatorial region were proposed by \citet{1993ApJ...409..429B} to explain the observed disks, but this is not likely the dominant mode for formation \citep{1996ApJ...472L.115O, 1995ApJ...440..308C}. Likewise, processes involving strong magnetic fields have been {raised} \citep[e.g.,][] {2002ApJ...578..951C}, but ultimately rejected \citep{2016ASPC..506..207W, 2018MNRAS.478.3049U}. To date, it seems that the most viable explanation can be found in non-radial pulsations (NRPs), as suggested by \citet{1986PASP...98...30O}. \citet{2016A&A...588A..56B} showed that {the combination of multiple NRP modes could remove energy and angular momentum deficiencies that limit outbursts}, and \citet{2022AJ....163..226L} illustrated that pulsations are very often organized in closely-spaced groups of frequencies. In a specific example, \citet{2020A&A...644A...9N} found that certain NRP modes allowed for the transfer of energy towards the surface layers of HD 49330, aiding in an outburst. Furthermore, \citet{2021MNRAS.508.2002R} demonstrated that a periodic phenomenon of disk building for HD 6226 was related to two closely-spaced NRP frequencies. {On the other hand, or perhaps in parallel, it has been suggested that Rossby waves may become mechanically excited and/or enhanced in visibility during mass ejection activity, perhaps in response to the outburst \citep{2018MNRAS.474.2774S, 2024bss..confE..76S}.} Other earlier works provide a theoretical framework for NRPs \citep[e.g.,][]{2009ApJ...701..396C}, and observational evidence of NRPs was presented for the star $\mu$ Centauri by \citet{1998ASPC..135..343R}.

Disk formation and dissipation for Be stars can aid our understanding of astrophysical disks, and thus observations are needed to constrain how the disks evolve. We can measure the changing disk through the strength of emission lines, which generally increase during disk growth, and decrease during dissipation. Paper I discusses the circularization and VDD modeling in detail, so we adopt that framework and focus our new analyses on multi-line EW and V/R evolution. 

\begin{table}[htbp]
\centering
\caption{Observations used in this work.}
\label{tab:obs_summary}
\begin{tabular}{l l c c c c l}
\hline\hline
\multicolumn{6}{c}{\textbf{Spectroscopic data}} \\ \hline 
Instrument & Date range (civil) & HJD\tablenotemark{a} & \# exposures & Typical exptime (s) & Typical S/N \\
\hline
CHIRON\tablenotemark{b}  & 2023 Jul 04 -- 2023 Aug 31 & 2460130 - 2460188 & 526 & 250 & $\sim$150 \\
NRES\tablenotemark{c}  & 2023 Jun 05 -- 2023 Sept 07 & 2460101 - 2460194 & 172 & 400--800 & $\sim$300 \\ \hline
\multicolumn{6}{c}{\textbf{Photometric (TESS) data}} \\ \hline 
Sector & Date range (civil) & HJD & \# points & Cadence (min) & Noise Floor\tablenotemark{d} {(ppt)} \\ \hline
13 & 2019 Jun 19 -- 2019 Jul 17 & 2458653 - 2458682 & 1283 & 30 & 0.0054 \\
66--67 & 2023 Jun 02 -- 2023 Jul 29 & 2460097 - 2460155 & 22,027 & 3.33 & 0.0054 \\ 
93--94\tablenotemark{e} & 2025 Jun 03 -- 2025 Jul 25 & 2460829 - 2460882 & 20,417 & 3.33l
& 0.0054 \\
\hline
\end{tabular}
\tablenotemark{Additional Note: For reference, the outburst took place from HJD 2460152 - 2460166 for all \ion{He}{1} lines.}
\tablenotetext{a}{HJD = Heliocentric Julian Date}
\tablenotetext{b}{Resolving Power $\sim$ 80,000}
\tablenotetext{c}{Resolving Power $\sim$ 53,000}
\tablenotetext{c}{The noise floor is hard to measure for each data set, but this represents a reasonable value for the combined light curve.}
\tablenotetext{d}{Photometric data not previously analyzed in Paper I.}

\end{table}

This {work studies} the southern star $\lambda$ Pav, a member of the Scorpio-Centaurus association, with spectral type B2IIIe \citep{10.1111/j.1365-2966.2006.10655.x}. Early works on $\lambda$ Pav \citep[e.g.,][]{1949Obs....69...30G,1950ApJ...111..663S} described emission patterns. Later, \citet{1975ApJS...29..137S} and \citet{1989ApJ...347.1082C} measured rotational velocities and found them to be either $\sim170$ km s$^{-1}$ or $\sim210$ km s$^{-1}$. Most recently, \citet{2016A&A...595A.132Z} measured a $v\sin(i)$ of 145 km s$^{-1}$, which we use later in this work. Since these earlier studies, the focus has been on studying the variability of different lines. \cite{2003A&A...411..229R} noted that $\lambda$ Pav has line profile variability caused by NRPs, but lacked enough data to determine a period. \citet{2011A&A...533A..75L} performed Fourier analysis on time series spectroscopy, finding four frequencies: $\nu_{1,L}$ = 0.17 $\pm$ 0.02, $\nu_{2,L}$ = 0.49 $\pm$ 0.05, $\nu_{3,L}$ = 0.82 $\pm$ 0.03, and $\nu_{4,L}$ = 1.63 $\pm$ 0.04 cycles d$^{-1}$, which were interpreted as g-mode pulsation ($v_{4,L}$), orbiting circumstellar material ($\nu_{3,L}$), and likely aliases (i.e. without a physical explanation; $\nu_{1,L}$ and $\nu_{2,L}$). Paper I found that this was a system that began ejecting material to form a disk at the end of TESS sectors {66 and 67}, which we further explore here to include additional spectral lines and pulsations in this paper. Paper I {additionally reported that $\nu_{3,L}$ also appears} in the emission oscillation asymmetry cycles (the V/R frequency/period during the early stage of the outburst). But this frequency is also present in TESS photometry even when there is no disk material, as evident by the lack of any emission in contemporaneous spectroscopy. As these results are the main findings related to $\lambda$ Pav in Paper I, an in-depth analysis may be able to provide a better understanding of how the disk forms around this Be star, especially since this disk growth happened at a time when when there was no hint of any pre-existing circumstellar material.

We characterize the disk growth and subsequent decay, along with V/R variations in H$\alpha$, H$\beta$, \ion{He}{1} $\lambda$5876, $\lambda$6678, $\lambda$4921, and $\lambda$4713 in Section 2. We then present a frequency analysis of the TESS photometry in Section 3, allowing us to present the photometric oscillations of the star with LPVs before, during, and after the outburst and intra-night rapid LPVs in Section~4. We discuss these findings in Section~5 and conclude in Section~6.

\section{A Well-Observed Disk Outburst in 2023} \label{sec:obs_outburst}

\begin{figure}[t!]
    \centering
    \includegraphics[width=1\linewidth]{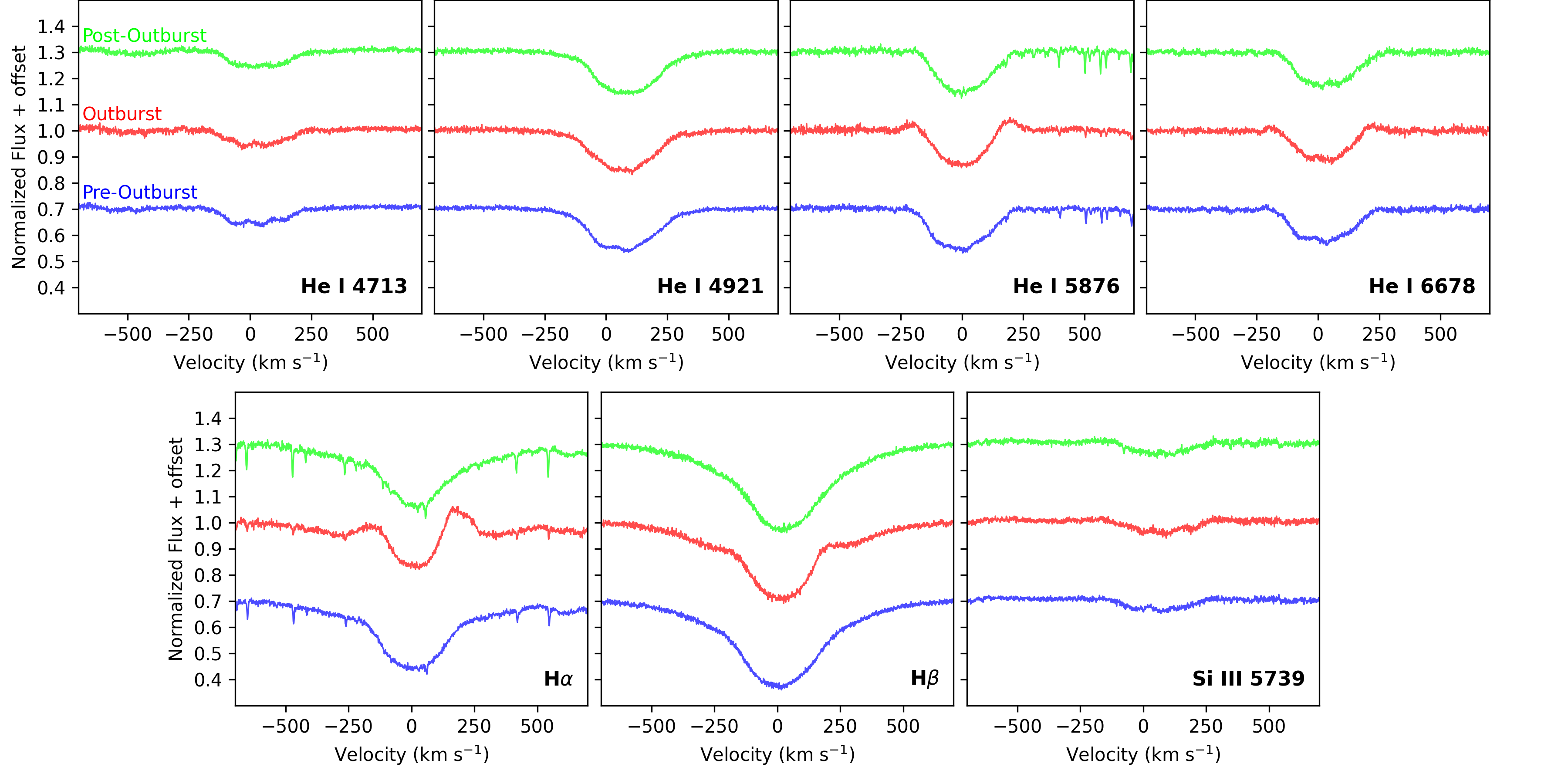}
    \includegraphics[width=0.45\linewidth]{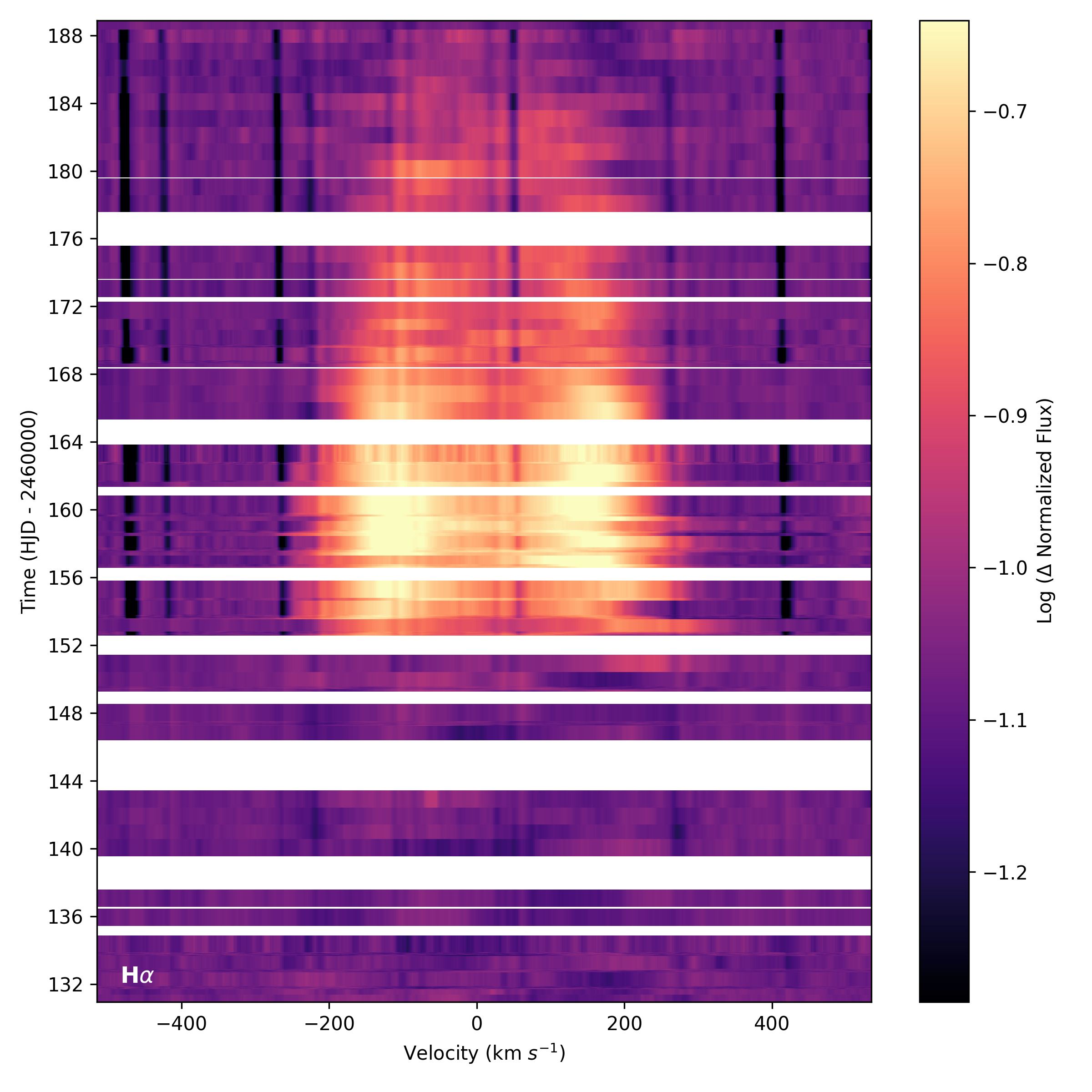}
    \includegraphics[width=0.45\linewidth]{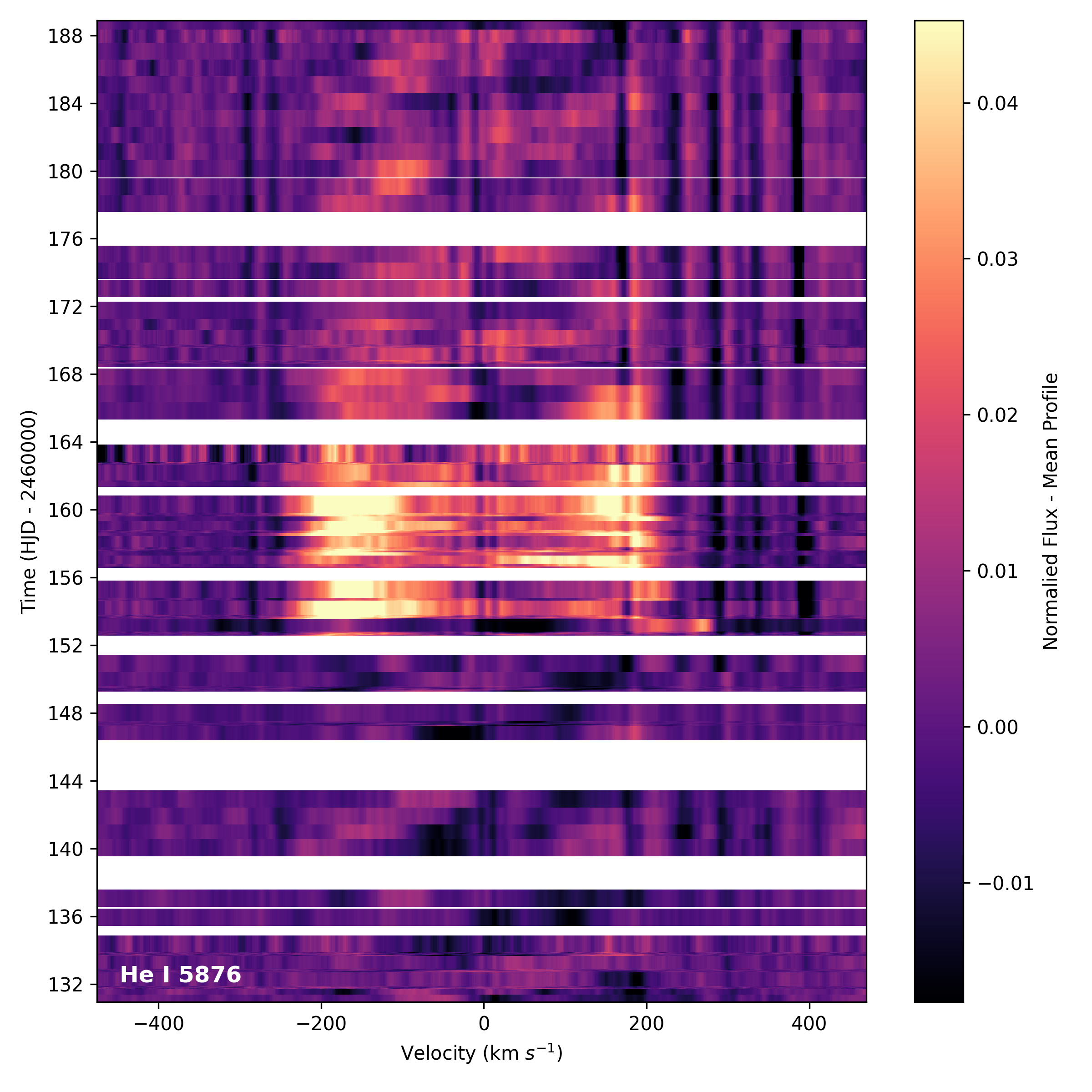}
    \caption{Seven spectral lines studied in this paper are displayed on the top. Included are representative spectra from 2023 July 7 {(HJD 2460133)} prior to the outburst, 2023 July 31 {(HJD 2460157)} during the outburst{, right after the end of TESS sector 67}, and 2023 August 26 {(HJD 2460183)} after the outburst. Large variations are seen in the hydrogen and stronger helium lines, while weak changes are seen in lines like \ion{He}{1} $\lambda$4713. No strong {circumstellar} changes were seen in the \ion{Si}{3} $\lambda$5739 line. In the bottom two panels are dynamical spectra of H$\alpha$ (left) and He I 5876 (right) presenting the entire time series. In these plots, we have combined all the spectra taken prior to the outburst to an average profile that was subtracted from all spectra, thus showing two main features: the emission from the disk during the outburst and, in the case of \ion{He}{1} 5876 the line profile variations, especially on the dominant $\sim6$ d period, are seen as darker perturbations moving from blue to red.} 
    \label{fig:spec_line_profiles}
\end{figure}

It is worth understanding the original observing strategy for the spectroscopic data collection in order to put our results in context.
Our observational campaign began with the hope to understand the spectroscopic and photometric properties of $\lambda$ Pav's NRPs as the star showed no sign of a disk in the early times we observed the star with NRES during TESS sector 66. In case of an outburst {during TESS sector 67}, we monitored the star and checked H$\alpha$ for emission with NRES, with additional intense monitoring for pulsational signatures and fast variations from CHIRON during the TESS observation window. For the majority of this time, no emission features were present. We then observed the sudden appearance of H$\alpha$ and H$\beta$ emission on 2023 {July 26 (HJD 2460152)}, not long before the end of the TESS observations {of $\lambda$ Pavonis for the year}. We changed our approach for observing the target, continuing the spectroscopic monitoring after the TESS observations and obtaining additional intense time-series to examine potential pulsational properties both during and after {the} outburst of disk activity in this Be star.

In the top panels of Fig.~\ref{fig:spec_line_profiles} are spectral line profiles for seven lines that were used in this study available in both the CHIRON and NRES spectra. We highlight spectra from before, during, and after the outburst. The bottom panels of Fig.~\ref{fig:spec_line_profiles} show the evolution of H$\alpha$ and \ion{He}{1} $\lambda$5876 lines in a dynamical representation and indicate that emission began around July 26, 2023 (HJD 2460152). There was a gap in observations around this date, so what we can say definitively is that there were no signs of emission before July 26, but that emission had started afterwards in all lines. This agrees with the EW measurements, which indicate that emission started concurrently in all lines with emission. By 2023 August 10 (HJD 2460166), emission levels had dissipated in all \ion{He}{1} lines, but Balmer emission levels had not dissipated to the continuum level until 2023 August 20 (HJD 2460177). The dynamical representation displays the fast buildup of the disk with the material in emission that is seen to move around the profile center, until the line is more symmetric as the disk circularized, and finally begins to dissipate. Furthermore, especially in the \ion{He}{1} profile, the difference profiles in Fig.~\ref{fig:spec_line_profiles} show clear $\sim6$ d blue-to-red {migrations}, which is most easily seen outside of the disk outburst. 

\begin{figure}[ht!]
    \centering
    \includegraphics[width=0.95\linewidth]{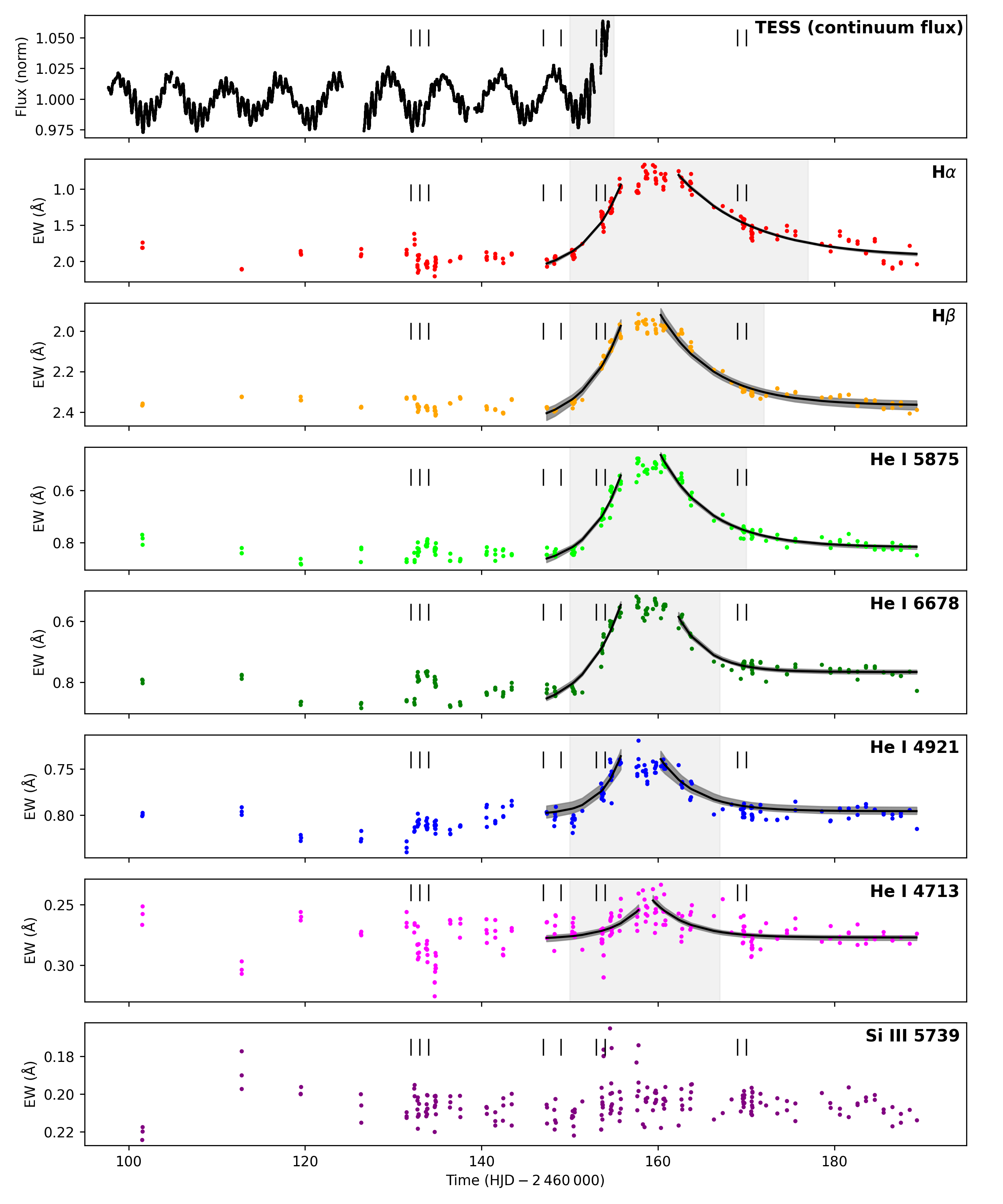}
    \caption{TESS light curve for sectors 66 and 67, which ends on day 155, and the EW measurements for the combined CHIRON and NRES data during the summer of 2023 for the seven lines studied. The gray regions indicate the time during the outburst -- when emission features are visually present in the spectra. \ion{H}{1} lines in particular take longer to decay, so the time during the outburst is longer than other lines. We then overplot an exponential growth and decay model for six of the seven spectral lines calculated with the MCMC sampler, where the black line is the maximum likelihood model and the gray region is the 1$\sigma$ posterior spread. Since \ion{Si}{3} $\lambda5739$ shows no appreciable emission, it is not modeled. The vertical dashes indicate times with intense spectroscopic coverage.}
    \label{fig:EW}
\end{figure}

In addition to the visual inspection, we measured the EW for the spectral lines displayed in Fig.~\ref{fig:spec_line_profiles}. In Fig.~\ref{fig:EW}, we compare EW measurements for H$\alpha$, H$\beta$, \ion{He}{1} $\lambda$4713, \ion{He}{1} $\lambda$4921, \ion{He}{1} $\lambda$5875, \ion{He}{1} $\lambda$6678, and \ion{Si}{3} $\lambda$5739 with the TESS flux where we indicate the timing of the outburst for each set of measurements in gray. The Balmer and \ion{He}{1} lines rise sharply in their emission beginning around {July 26 (HJD 2460152)}, corresponding to the start of emission and the {increase} in flux in the TESS data. Then, EW values of the \ion{He}{1} lines gradually return to pre-outburst levels by {August 10 (HJD 2460166)}, consistent with the dissipation of the disk, and the end of the disk growth. The higher excitation \ion{He}{1} lines at 4713 and 4921 \AA\ returned to their pre-outburst level faster than the \ion{He}{1} 5875 and 6678 lines. Outside of the outburst window, the EW curves are stable, confirming that $\lambda$ Pav underwent the entire outburst event completely contained within a roughly three-week time frame. However, we note that there may be a slight rise in the level of the line emission following the outburst compared to prior to the outburst.

With the data we have available to us, we can quantify the decay of the disk. The VDD model is described by \citet{2006ApJ...639.1081C,2008ApJ...684.1374C}, and has had success in many cases \citep[e.g.,][and references therein]{2021MNRAS.508.2002R}.  \citet{2016ASPC..506..157R} found a general form for the disk dissipation given by
\begin{equation}
m(t) = m_0 - (\Delta_{\rm mag}) e^{\frac{-\xi_{\rm band}(t - t_0)}{\tau}}.
\end{equation}
In this equation, $m(t)$ is the time-dependent magnitude, $m_0$ is the quiescent magnitude, $\Delta_{\rm mag}$ is the amplitude of the decay, $\frac{\xi_{\rm band}}{\tau}$ dictates the timescale of the decay of the light curve, and $t_0$ is a zero-point in time. {These parameters are given in equation 5 in \citet{2016ASPC..506..157R}, but characterized with filters for Johnson $BVRI$ photometry rather than the generalized bandpass we used here. } The VDD model has been used for the decay of the disk, but a similar model for growth has not yet been established for short outbursts like the one we observed for $\lambda$ Pav, though \citet{2018MNRAS.476.3555R} have developed models for longer-duration outbursts. This exponential decay has been used previously to describe the decay of the disk signatures in spectral lines \citep[e.g.,][]{2021MNRAS.508.2002R}, so we use a form that handles our EW values, namely
\begin{equation}
EW(t) 
= EW_{0} + \bigl(EW(\rm peak) - EW_{0}\bigr)e^{\frac{-(t - t_{\rm peak})}{\Delta t}}.
\end{equation}
Here $EW_{0}$ is the quiescent EW, $EW(\rm peak)$ is the maximum EW reached at the time of peak emission, $t_{\rm peak}$, and $\Delta t$ is the time for the curve to fully decay. 

We can fit the hydrogen and \ion{He}{1} lines with an MCMC sampler procedure using Eq.~2 as our model, along with the data near the peak of the outburst and onwards to the end of the 2023 spectroscopic time-series data. We developed the model with the {\tt emcee} code described by \citet{2013PASP..125..306F}, which ran with 256 walkers, and after running a 100‑step burn‑in, was followed by 1,024 iterations. Our posterior spread of models was examined with the {\tt corner} package in python \citep{corner} indicating that high-quality fits were made, with an example shown at the end of this paper in Fig.~\ref{fig:corner}. The maximum likelihood model is also calculated via the {\tt lnprob()} function and is the posterior sample with the largest recorded log-likelihood after burn-in. The \ion{Si}{3} 5739 line from Fig.~\ref{fig:EW} shows no detectable emission, so it may not be useful for understanding the disk, but various lines that contain no emission will be useful for understanding pulsational behavior. {$1\sigma$ posterior spreads are overplotted with their fits in Fig.~\ref{fig:EW}}. From the fits, we include in Table~\ref{tab:fit_params} {the parameters and} the errors given in the Monte Carlo sampling output. We see that the hydrogen lines decay over a longer time frame ($\Delta t$=8.5 days for H$\alpha$ and $\Delta t$=6.4 days for H$\beta$) than the helium lines ($\Delta t \approx$ 2.4--5.6 days). With a positive exponent rather than a negative, we fit the rise of the disk with results in Table \ref{tab:fit_params} and Fig.~\ref{fig:EW}, although this is not as physically motivated as the decay parameters. Within Table \ref{tab:fit_params}, not all errors are symmetric, and may be underestimated \citep[see, e.g.,][]{2021MNRAS.502.5038N}. While these errors are slightly increased for parameters that had an error of 0, the posterior spread in the models in Fig.~\ref{fig:EW} {implies} that our fits are realistic.

\begin{table}[ht!]
\centering
\caption{Fit parameters for disk growth, circularization, and decay for each spectral line.}
\label{tab:fit_params}
\begin{tabular}{lcccccc}
\hline
Parameter & H$\alpha$ & H$\beta$ & He\,I\,5876 & He\,I\,6678 & He\,I\,4921 & He\,I\,4713 \\
\hline

\multicolumn{7}{c}{\textbf{Disk growth parameters}} \\
\addlinespace
$EW_{0}$ (\AA) \tablenotemark{a}                    & $2.20 \pm 0.05$ & $2.47 \pm 0.04$ & $0.90^{+0.03}_{-0.02}$ & $0.90 \pm 0.02$ & $0.80 \pm 0.01$ & $0.28 \pm 0.01$ \\
$EW_{\mathrm{peak}}$ (\AA)         & $0.65 \pm 0.05$ & $1.80 \pm 0.05$ & $0.49 \pm 0.01$ & $0.51 \pm 0.02$ & $0.71 \pm 0.01$ & $0.24 \pm 0.01$ \\
$t_{\mathrm{peak}}$ (HJD$-2\,460\,000$)   & $156.67^{+0.18}_{-0.17}$ & $157.09^{+0.49}_{-0.44}$ & $156.34^{+0.23}_{-0.21}$ & $156.26^{+0.26}_{-0.25}$ & $156.70^{+0.60}_{-0.52}$ & $158.79^{+0.65}_{-0.68}$ \\
$\Delta t$ (days)                         & $4.15^{+0.39}_{-0.35}$ & $4.30^{+0.45}_{-0.44}$ & $3.87^{+0.66}_{-0.54}$ & $4.35^{+0.43}_{-0.41}$ & $2.55^{+0.71}_{-0.63}$ & $3.32^{+0.47}_{-0.42}$ \\

\hline
\multicolumn{7}{c}{\textbf{Circularization parameters}} \\
\addlinespace
$\left(\rm{V/R}\right)_0$ & $0.98\pm0.01$ & $1.02\pm0.01$ & $1.04\pm0.01$ & $1.04\pm0.01$ & $1.01\pm0.01$ & $1.06\pm0.01$ \\
$A$ (amplitude)                             & $0.35\pm0.01$ & $0.09\pm0.01$ & $0.35^{+0.02}_{-0.01}$ & $0.32 \pm 0.02$ & $0.10\pm0.01$ & $0.19\pm0.01$ \\
$P$ (days)                                  & $1.23\pm0.01$ & $1.25\pm0.01$ & $1.25\pm0.01$ & $1.25\pm0.01$ & $1.25\pm0.01$ & $1.27\pm0.01$ \\
$t_0$ (HJD$-2\,460\,000$)                   & $152.51\pm0.01$ & $152.48 \pm 0.03$ & $152.45\pm0.01$ & $152.43\pm0.01$ & $152.37\pm0.04$ & $152.26 \pm 0.02$ \\
$\tau$ (days)                               & $12.30^{+0.44}_{-0.46}$ & $11.88^{+0.97}_{-0.99}$ & $12.43^{+0.78}_{-0.74}$ & $11.54^{+0.76}_{-0.79}$ & $11.80^{+0.97}_{-0.99}$ & $11.79^{+0.94}_{-0.96}$ \\

\hline
\multicolumn{7}{c}{\textbf{Decay parameters}} \\
\addlinespace
$EW_{0}$ (\AA)                     & $1.95\pm0.03$   & $2.37\pm0.02$    & $0.82\pm0.01$   & $0.77\pm0.01$    & $0.79\pm0.01$   & $0.28\pm0.01$ \\
$EW_{\mathrm{peak}}$ (\AA)         & $0.65\pm0.01$   & $1.80\pm0.02$    & $0.49\pm0.01$   & $0.56\pm0.02$    & $0.71\pm0.01$   & $0.24\pm0.01$ \\
$t_{\mathrm{peak}}$ (HJD$-2\,460\,000$)   & $161.21^{+0.26}_{-0.28}$ & $158.75^{+0.61}_{-0.71}$ & $160.72 \pm 0.30$ & $161.82 \pm 34$ & $158.25^{+0.63}_{-0.64}$ & $159.79^{+0.53}_{-0.47}$ \\
$\Delta t$ (days)                         & $8.51^{+0.64}_{-0.59}$  & $6.29^{+0.80}_{-0.77}$  & $5.67^{+0.50}_{-0.47}$  & $3.39^{+0.42}_{-0.40}$  & $4.20^{+0.81}_{-0.75}$  & $3.86^{+0.66}_{-0.64}$ \\

\hline
\end{tabular}
\tablenotetext{a}{We measured the EW of all lines in the velocity range of $\pm500$ km s$^{-1}$ from the rest wavelength.}
\end{table}

\begin{figure}[ht!]
    \centering
    \includegraphics[width=1\linewidth]{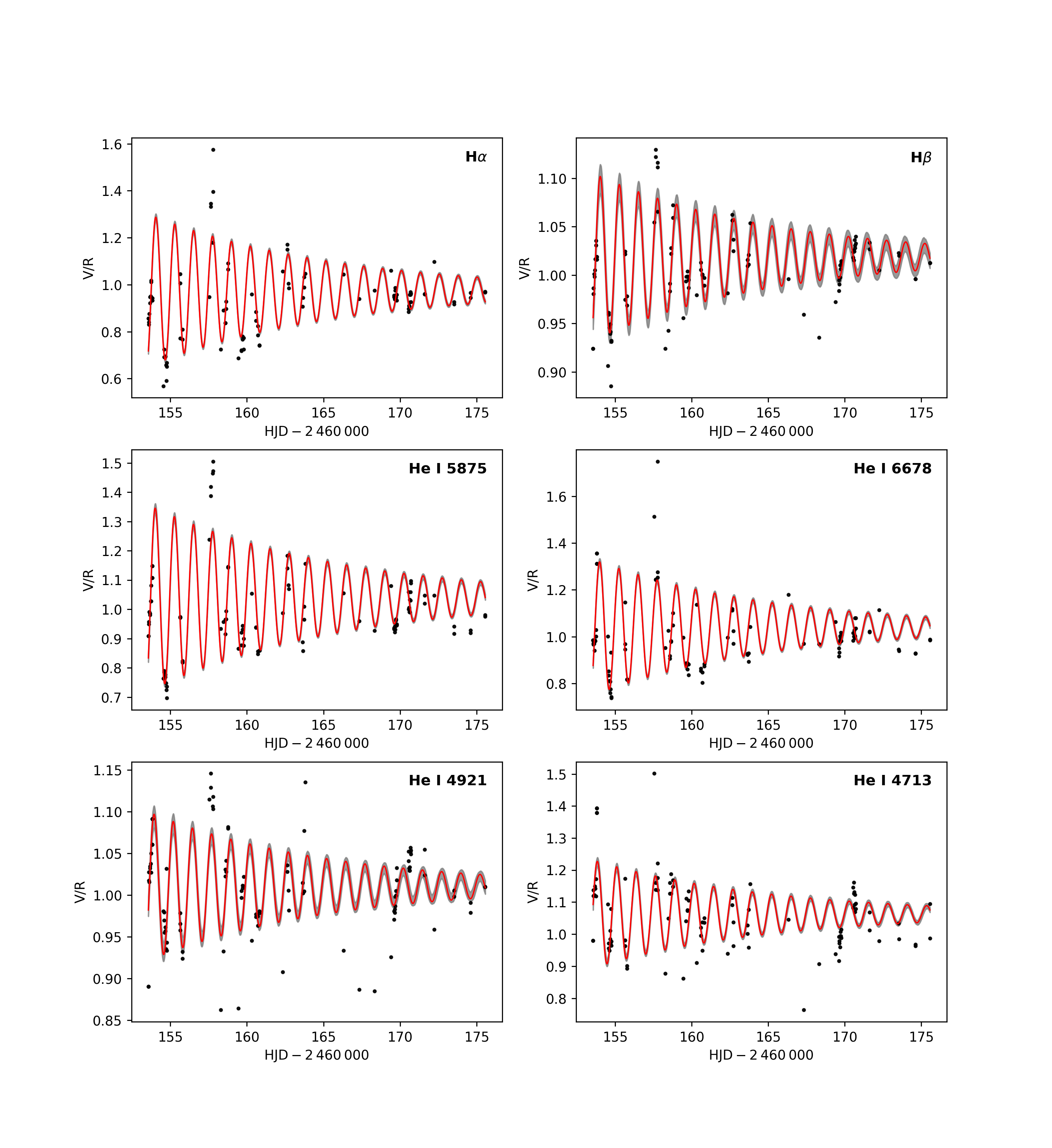}
    \caption{A fit to the V/R measurements in H$\alpha$ (top left), H$\beta$ (top right), \ion{He}{1} 5875 (middle left), \ion{He}{1} 6678 (middle right), \ion{He}{1} 4921 (bottom left) and \ion{He}{1} 4713 (bottom right) {using Eq.~4}. The gray region represents a $1\sigma$ posterior spread, and the red line is the maximum likelihood model. {The variation in amplitude line by line results in different y-axis scales. Also, the mean V/R is slightly above 1 in most lines (also included in Table~\ref{tab:fit_params}) although the V/R had an average of unity prior to the outburst. This small deviation could be instrumental or real in these measurements as data were taken with two different observatories and spectrographs. Additional offsets besides heliocentric corrections were typically less than $\sim$10 km s$^{-1}$}.}
    \label{fig:VRfit}
\end{figure}

High-resolution spectroscopy also allows us to explore the shape of the outburst and the asymmetry of the emission with time, which has historically been measured with a violet to red (V/R) ratio for the emission. This is not a well-defined measurement in the literature, with measurements including a direct measurement of the peak heights, peak heights adjusted for a photospheric correction, or the EW of the two sides of the spectral line. In this analysis, the V/R ratio is computed as $V/R = \frac{\rm{{EW_{V}}}}{\rm{EW_{R}}}$ by integrating the EW from the line center out to beyond the points of emission, then taking the ratio of the blue ($\lambda$ $<$ $\lambda_{{c}}$) to red ($\lambda$ $>$ $\lambda_{{c}}$) halves, where $\lambda_c$ denotes the line center. For example, the H$\alpha$ line center was -15 km s$^{-1}$ from the rest velocity associated with a 6562.81 \AA \, wavelength, and the width used to capture the emission features was 400 km s$^{-1}$. After measuring these ratios for each line that shows disk emission, we explored modeling these variations with time. 

In Paper I's recent analysis of many H$\alpha$ and photometric outbursts recorded with TESS, it was demonstrated that V/R variations in Be stars can be described with a time–dependent sinusoid of the form
\begin{equation}
y(t) = A + \bigl(B + Ct\bigr)\sin\biggl(2\pi\frac{t - F}{D + Et}\biggr),
\end{equation}
where A is the quiescent V/R value, B is the initial amplitude, C is the rate of change of amplitude with time, D is the period, E is the rate of change in the period, and F is a phase offset. This form allows for both the amplitude and period to vary, but our spectroscopic data for $\lambda$ Pav seems to illustrate that the V/R oscillations themselves fade in strength following the outburst. We modeled the V/R variations with a sinusoid modulated by an exponential decay as our dense time-series seemed to support this over the linear decay model, namely 
\begin{equation}
\rm{V/R}(t) = \rm{V/R}_0 + A\Bigl(e^{\frac{-(t - t_{{peak}})}{\tau}}\Bigr)
\sin\Biggl(\frac{2\pi(t - t_{{peak}})}{P}\Biggr),
\end{equation}
where V/R$_0$ is the quiescent V/R level, A the initial oscillation amplitude, P the period, $t_{{peak}}$ is the phase zero point, and $\tau$ the time of the amplitude decay.  

We fit the resulting V/R time series using an {\tt emcee} sampler \citep{2013PASP..125..306F} with 240 walkers, running a 100‑step burn‑in followed by 1,024 iterations. Fig.~\ref{fig:VRfit} displays the V/R data and our maximum‑likelihood fit with the resulting posterior spread. We also include the {\tt corner} plots online with the plots from the decay sampler fits. The fits in Fig.~\ref{fig:VRfit} illustrate how well constrained the posterior distributions are for V/R$_0$, A, P, $t_0$, and $\tau$ for all of the spectral lines examined, with the parameters included in Table \ref{tab:fit_params}. Although the form used in this work differs slightly from Paper I in how it allows amplitude (and period) to vary in that we use an exponential decay compared to a linear decay, the inclusion of $\tau$ ensures a physically meaningful decay in line with the disk’s evolution. The period (1.25 $\pm$ 0.01 d) of the sinusoidal term agrees with the H$\alpha$ measurements of Paper I {which} reported a frequency of 0.81 c d$^{-1}$. These fits and the associated error estimates are given in Table \ref{tab:fit_params}.

\section{Periodic Behavior}
\begin{figure}[t!]
    \centering
    \includegraphics[width=0.95\linewidth]{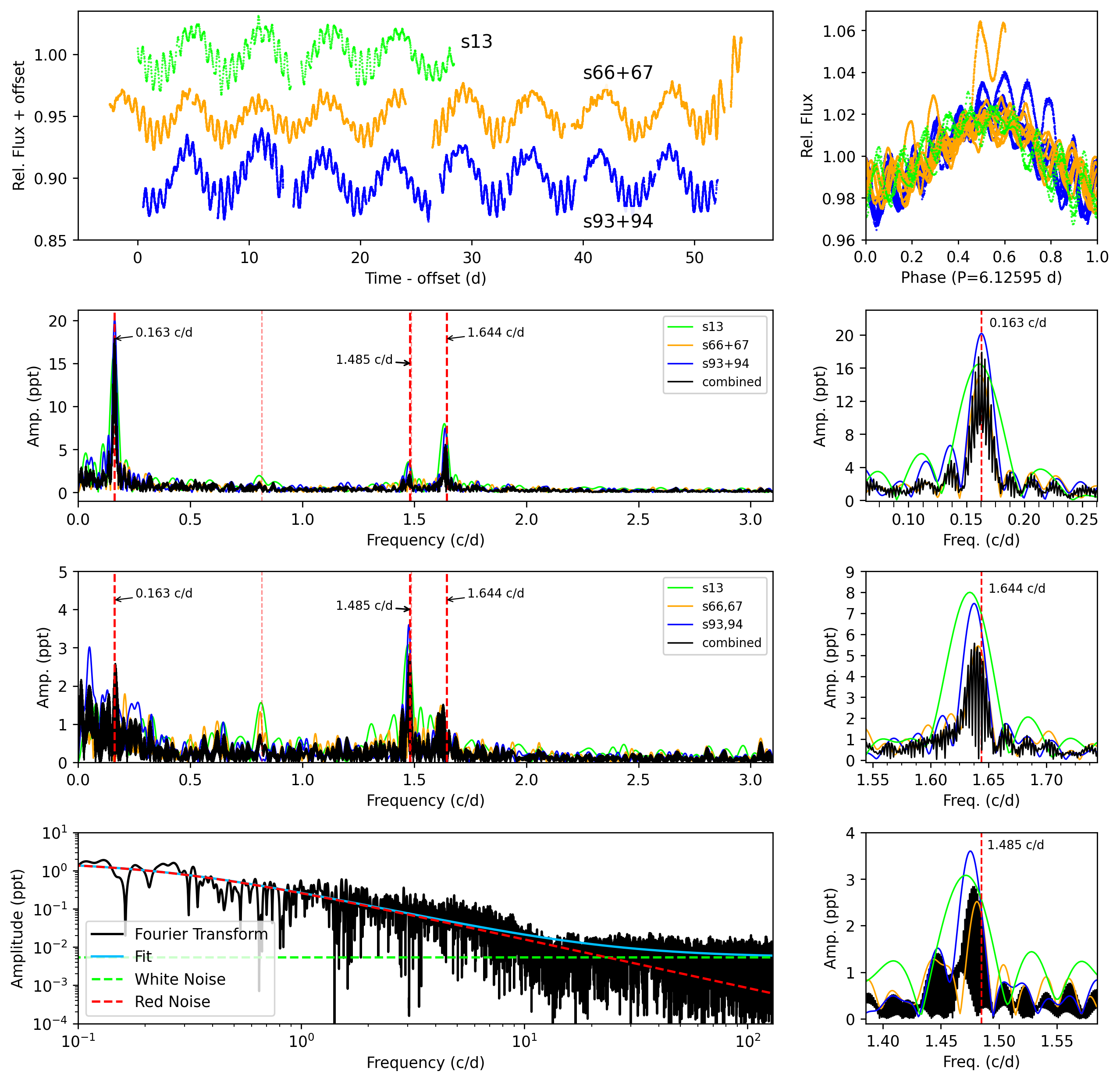}
    \caption{{In the left column are the TESS light curves for each sector of observation in raw flux (top), the Lomb–Scargle periodograms of each TESS light curve {calculated with {\tt Period04} \citep{2004IAUS..224..786L}} (top middle), the Fourier transform after removing the two {highest amplitude} frequencies, which displays the third {highest amplitude} frequency at 1.485 d$^{-1}$ more clearly (bottom middle), and the Fourier transform with all frequencies removed and then fit (bottom, see text for further details). On the right are panels that show the phased TESS light curves (top), zoom-ins of the three dominant frequencies: 0.163 d$^{-1}$ (top middle, from top middle row), 1.644 d$^{-1}$ (bottom middle, from top middle row), and 1.485 d$^{-1}$ (bottom, from bottom middle row). The red dashed lines mark the three strongest signals and an additional red dotted line in the left panels highlights the 0.82 d$^{-1}$ spectroscopic frequency from \citet{2011A&A...533A..75L}.}}
     \label{fig:FT}
\end{figure}

Thus far, our discussion of the TESS light curve taken contemporaneously with our spectroscopy has focused on the outburst in the last few days of the TESS sector 67 data (Fig.~\ref{fig:EW}). The behavior seen in all currently available TESS light curves (sectors 13, 66, 67, 93, and 94) is remarkably similar. We used the program {\tt Period04} \citep{2004IAUS..224..786L} to conduct a Fourier Analysis of the TESS data in order to quantify the photometric signals, analyzing the data in three sections (sector 13, 66+67, and 93+94) as shown in Fig.~\ref{fig:FT}. We performed the Fourier analysis in this manner as the adjacent sectors (66+67 and 93+94) only include a short gap in the data collection. In all cases, the dominant feature appears at a frequency of about $\nu_1 = 0.163\pm0.010$ d$^{-1}$ with an amplitude of about 20 parts per thousand (ppt). The next two strongest signals are found at $\nu_2 =1.644\pm0.010$ d$^{-1}$ (amplitude of about 6 ppt) and $\nu_3 =1.485\pm0.010$ d$^{-1}$ (amplitude of about 3 ppt). These three signals have an approximate numerical relation where the difference between the two higher-frequency signals corresponds closely to that of the lowest frequency signal (i.e. $\nu_1 = \nu_2 - \nu_3$). We used the dominant frequency, corresponding to a period of 6.13 d, to phase the five sectors of TESS photometry in Fig.~\ref{fig:FT}, which is remarkably stable across the years of data. {Despite this persistence, $\nu_1$ is notably lower in amplitude during sectors 66 and 67 compared to the other sectors ($\sim$1 ppt less than sector 13; $\sim$5 ppt less than sectors 93 and 94), even though the amplitude appears to increase as the outburst begins at the end of sector 67. The other 2 frequencies from sectors 66 and 67 appear lower in amplitude than their respective frequencies in the other 2 sector groups, as well.}
Pre-whitening against the two strongest peaks reveals several additional signals at low amplitudes (third row panel of Fig.~\ref{fig:FT}), generally organized in closely spaced groups, similar to the majority of other Be stars \citep{2022AJ....163..226L}. 

Viewing the frequency spectrum after removing all periodic variations shows a red noise profile characteristic of stochastic low-frequency variability (bottom panel of Fig.~\ref{fig:FT}) as seen in virtually all luminous OB stars \citep[e.g.,][]{2020A&A...640A..36B}. 
We fit the frequency spectrum with the equation
\begin{equation}
    \alpha_{\nu} = \frac{\alpha_0}{1+(\frac{\nu}{\nu_{char}})^{\gamma}} + C_w.
\end{equation}
This equation is based on the characterization of the Fourier properties of the Sun as described by \citet{1985ESASP.235..199H}, \citet{2014A&A...570A..41K} and others. Here, the amplitude of the Fourier transform $\alpha_\nu$ is a function of frequency $\nu$. $\alpha_0$ represents the amplitude of the semi-Lorentzian fit at a frequency value of zero, $\nu_{char}$ denotes a characteristic frequency, $\gamma$ is the logarithmic amplitude gradient, and $C_w$ is the white noise of the data. From this fit, we derive $\alpha_0 = 1.7964 \pm 0.0043$ ppt, $\nu_{char} = 0.2466 \pm 0.0012$ d$^{-1}$, $\gamma = 1.2741 \pm 0.0030$, and $C_w = 0.00538 \pm 0.00007$ {ppt}. 

Included in Fig.~\ref{fig:FT} are vertical dashed lines marking three frequencies found in {the TESS photometry} reported by \citet[][$\nu_{1,L}$ = 0.17 $\pm$ 0.02, $\nu_{3,L}$ = 0.82 $\pm$ 0.03, and $\nu_{4,L}$ = 1.63 $\pm$ 0.04 cycles d$^{-1}$]{2011A&A...533A..75L}. A fourth line marks the 1.485 cycles d$^{-1}$ signal, which is a new frequency with this work. Their lowest frequency signal, $\nu_{1,L}$, is confirmed by TESS to be genuine, and not an alias as \citet{2011A&A...533A..75L} presumed, while their $\nu_{4,L}$ is also obvious in the TESS data. 
There is a signal at their  $\nu_{3,L}$ in two of the three sections of TESS data. While there is no hint of their $\nu_{2,L}$ 0.49 $\pm$ 0.05 cycles d$^{-1}$ in the TESS photometry, there is a signal at 1.485 cycles d$^{-1}$, so it is possible that their $\nu_{2,L}$ is a one day alias.
{\citet{2011A&A...533A..75L} remark that their $\nu_{4,L}$ could be the first harmonic of $\nu_{3,L}$. The corresponding frequencies of these two signals in TESS could support this interpretation, which would imply that the physical process producing the fundamental frequency, $\nu_{3,L}$, manifests in non-sinusoidal flux variations and non-sinusoidal line profile variations. On the other hand, given the relatively low frequency resolution of both the photometric and spectroscopic datasets of \citet{2011A&A...533A..75L}, this may simply be a numerical coincidence. At any rate, it is apparent that the photometric amplitudes of $\nu_{3,L}$ and $\nu_{4,L}$ are not tightly coupled -- $\nu_{3,L}$ is not detected in the final section of TESS photometry while $\nu_{4,L}$ retains a significant amplitude. While \citet{2011A&A...533A..75L} suggested that $\nu_{3,L}$ may be a circumstellar frequency where LPVs are caused by orbiting material, the fact that this signal is detected in TESS in the absence of any emission (e.g. throughout the entirety of sectors 66+67 prior to the onset of mass ejection) suggests that at least the photometric signal at this frequency is not purely circumstellar. }

It should be noted that signals detected in line profile variations may not necessarily correspond to readily detectable photometric signals (and vice-versa). {Furthermore,} the $\nu_{3,L}$ frequency reported by \citet{2011A&A...533A..75L} is remarkably similar to the frequency we derive for the {circumstellar} V/R oscillations, and seems to be marginally above the noise in the Fourier spectrum of the pre-whitened light curve for sectors 13 and 66+67 \citep[i.e. with the $\nu_{1,L}$ and $\nu_{4,L}$ signals from][removed]{10.1111/j.1365-2966.2006.10655.x}. It is somewhat difficult to see how an orbital or rotational timescale for the disk can also be present throughout entire sectors of TESS data with an outburst only being present over a few days of one of these sectors. 

All of the observed periodic signals in TESS are presumed to be photospheric in origin. The last $\sim$3 days of sector 67 (where the disk was starting to build up) were omitted from the frequency analysis, so that all signals from sectors 66+67 in Fig.~\ref{fig:FT} were recorded during a disk-less state. There was very weak emission present during sector 93, which had dissipated to be no longer detectable throughout sector 94. That is, no mass ejections occurred during these last two sectors. We lack any spectroscopic coverage of TESS sector 13, but there is no evidence of mass ejection in the sector 13 light curve. {That is, {the end of sector 67 has} a sharp rise in flux in Fig.~\ref{fig:FT}, but this is not present in the sector 13 flux.
}

\section{Line Profile Variations Behavior Before, During, and After the Outburst}

\begin{figure*}[ht!]
\centering
\includegraphics[width=\textwidth,height=17cm]{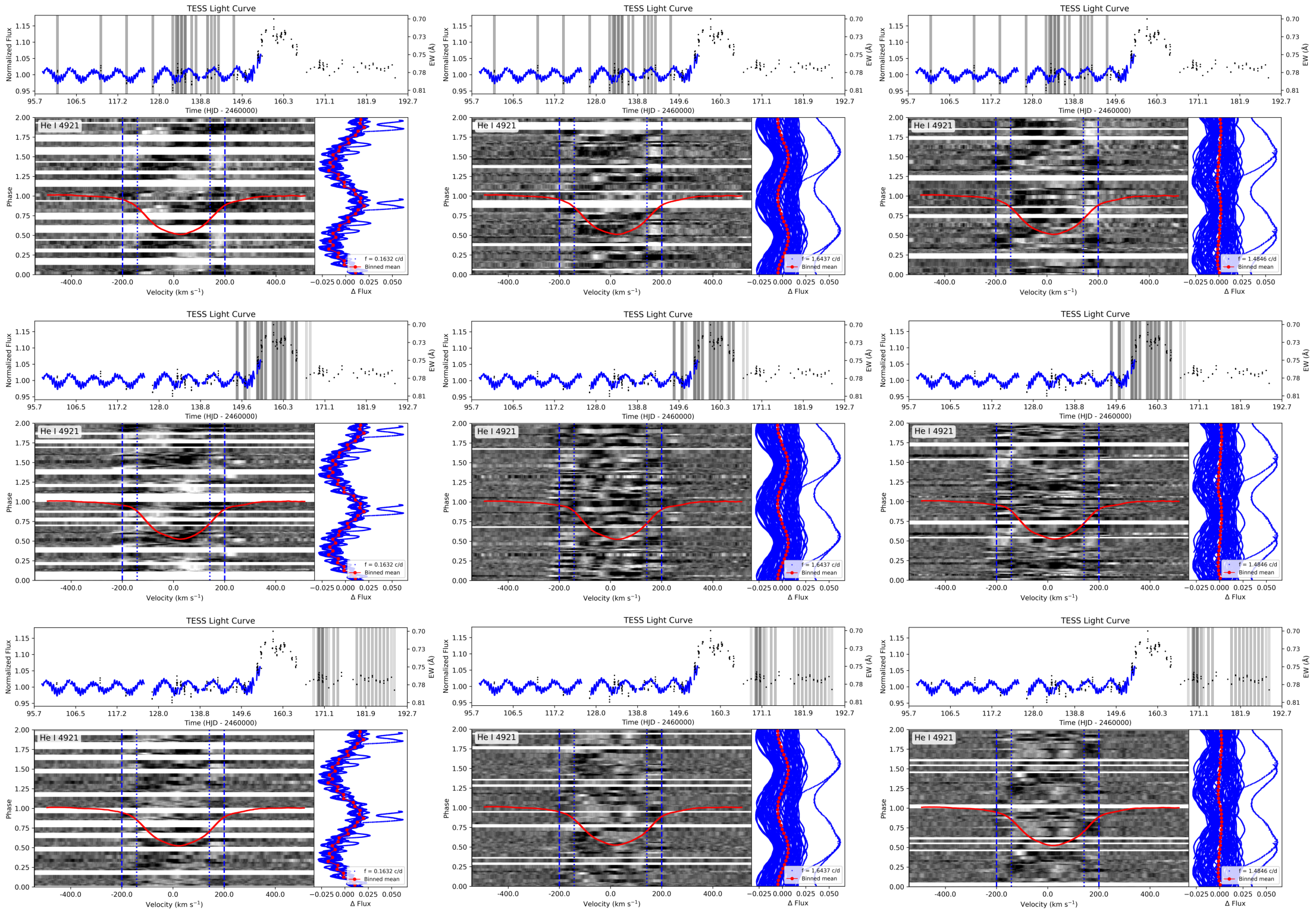}
    \caption{Spectroscopic data for \ion{He}{1} 4921 as related to the three dominant photometric frequencies derived by TESS, with the left column depicting the 0.163 d$^{-1}$ frequency, the middle column contains the 1.644 d$^{-1}$ frequency and the right lays out the 1.485 d$^{-1}$ frequency. The top row displays results from before the outburst, the middle row during the outburst {(July 26 - Aug. 10)}, the final row is after the outburst. For each gray scale depiction of the spectra, we have subtracted the average line profile in red. There are two sets of vertical blue lines, one denotes the $v\sin i$ of $\pm$ 145 km s$^{-1}$ recorded by \citet{2016A&A...595A.132Z} that we use as a reference; the other set simply corresponds with the visual extent of the spectral lines during the non-out-bursting phase. The top panel of each sub plot displays the TESS light curve and a scaled EW variability curve, with the {gray bars marking the spectra used in the gray scale images}. To the right of each gray scale is the phased TESS light curve for that frequency from sectors 66 and 67.}
    \label{fig:slow_LPVs}
\end{figure*}

With {two pulsation frequencies and one difference frequency} from TESS {($\nu_1$ = 0.163 d$^{-1}$, $\nu_2$ = 1.644 d$^{-1}$, $\nu_3$ = 1.485 d$^{-1}$)}, we phased line‑profile maps to examine the behavior of $\lambda$ Pav before, during, and after the outburst. In Fig.~\ref{fig:spec_line_profiles}, we see a slow, $\sim6$ d motion in the residual line profiles, but a better demonstration of the line profile variations seen in \ion{He}{1} 4921 before, during, and after the outburst is in Fig.~\ref{fig:slow_LPVs} {for the three prominent signals found in the previous section}. 

Each subplot in Fig.~\ref{fig:slow_LPVs} has a combination of the TESS light curve and scaled EW values to show the outburst for \ion{He}{1} $\lambda$4921, with times indicated in gray that coincide with the spectral data used. Then, we include a residual plot for the spectral line as a function of phase calculated with the three prominent TESS frequencies. Lastly, each gray scale residual of the spectral lines is compared to the phased TESS light curve from sectors 66 and 67 when our spectra were obtained.

The strongest photometric signal was $\nu_1 = 0.163$ d$^{-1}$, with an amplitude of about 20 ppt {from the Fourier analysis of the entire light curve} {(left column of Fig.~\ref{fig:slow_LPVs})}. The spectroscopic signal appears clearly in each spectral line, with a general trend of an absorption excess starting on the blue side of the line and progressing to the red over the 6.13 d period. The behavior is the same in all three subsets of the data, taken before, during, and after the disk outburst. {These figures do contain gaps in phase, which is not unusual for such a small frequency, but this does make the pattern within the LPVs harder to see.} One thing to note in the \ion{He}{1} and \ion{H}{1} lines is that the pulsational behavior 
{extends beyond the literature value of $v \sin i = \pm145$ km s$^{-1}$, out to approximately $\pm200$ km s$^{-1}$, marked with vertical dashed lines. There may be additional line broadening mechanisms besides rotation, and/or it is possible that the $v \sin i$ value from \citet{2016A&A...595A.132Z} is underestimated.}
During the outburst, activity and variability can be seen beyond the lines at $\pm 200$ km s$^{-1}$, indicating that there must be some disk material present {that impacts the observed LPVs as they should not extend to super-rotational velocities}.

\begin{figure*}[t!]
    \centering
    \includegraphics[width=0.62\linewidth]{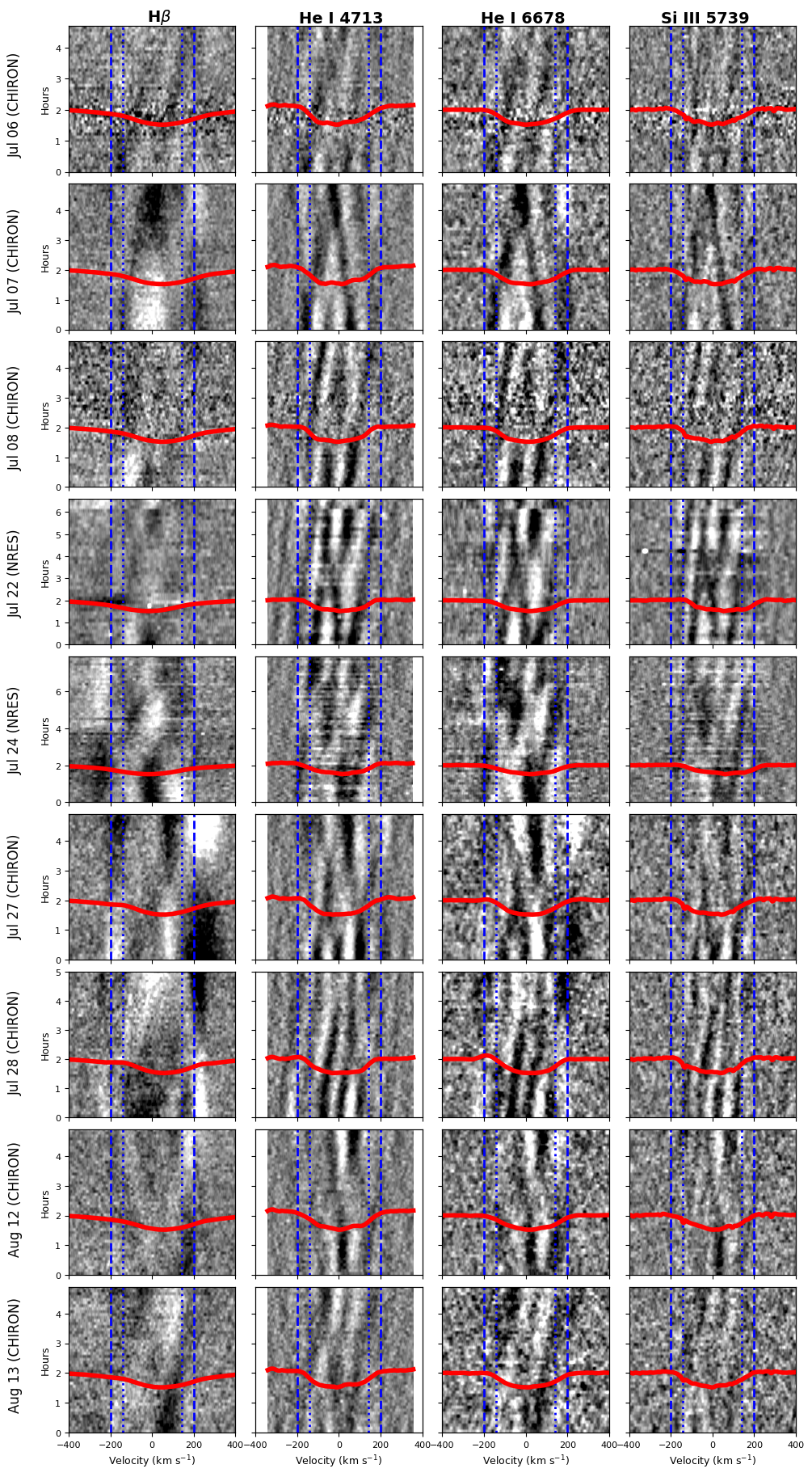}
    \caption{{Dynamical spectra} for H$\beta$ (left), \ion{He}{1} 4713 (second column), \ion{He}{1} 6678 (third column), and \ion{Si}{3} 5739 (right). Each row is one night showing short time-scale variations that vary from line-to-line, night-to-night. The H$\alpha$ variations are similar to those of H$\beta$, \ion{He}{1} 5876 similar to \ion{He}{1} 6678 variations, and the \ion{He}{1} 4921 variations similar to those of \ion{He}{1} 4713.
    \label{fig:fast_vary}} 
\end{figure*}

{The second ($\nu_2 =1.644$ d$^{-1}$) and third ($\nu_3 =1.485$ d$^{-1}$) {highest amplitude} frequencies in TESS are more challenging to recover in our spectroscopic data, especially when analyzing sub-sections of the dataset. This is likely due to a number of factors. First, the slow $\nu_1 = 0.163$ d$^{-1}$ signal dominates the periodic LPVs. Even in the \ion{He}{1} lines analyzed here, the net changes in the EW from circumstellar material have a relatively high amplitude compared to more subtle pulsational LPVs. The signal $\nu_2 =1.644$ d$^{-1}$ is about one d$^{-1}$ greater than the $\nu_1 = 0.163$ d$^{-1}$ signal, which likely causes problems with aliasing. Finally, the fast LPVs discussed in Sec.~\ref{sec:fastLPVs} are also relatively strong and contribute heavily to the `noise' when searching for LPV patterns at $\nu_2$ and $\nu_3$. }

After sub-dividing the spectroscopic data, the only hint of $\nu_2$ is seen in the pre-outburst data for \ion{He}{1} 4921 {(top-middle panel, Fig.~\ref{fig:slow_LPVs})}. Although marginally detected, the LPV pattern associated with $\nu_2$ is similar to that reported in \citet[][their Fig. 5]{2011A&A...533A..75L}, with the alternating light/dark patterns moving across the line profile from blue to red twice per phase. Due at least in part to the issues mentioned in the previous paragraph, $\nu_2$ is not recovered in the data strings during and after the outburst, and $\nu_3$ is not apparent in any data sub-strings.

Our {observing campaign} allowed us to obtain some very dense nights of spectroscopic data throughout the observational campaign. Thus we are also able {to} see if there are any short timescale variations over individual nights before, during, and after the outburst, although the spectroscopic data are not complete enough in time coverage for determination of periods. Fig.~\ref{fig:fast_vary} details the H$\beta$, \ion{He}{1} 4713, \ion{He}{1} 6678, and \ion{Si}{3} 5739 lines. Of the nine nights of dense spectroscopic coverage, we see fast-moving substructures in every time-series, likely related to high-order pulsations in the star that do not have a photometric period due to vanishingly smaller amplitudes {from cancellation effects across the stellar disk when the azimuthal order, m,} gets larger. As these are seen in every time-series, there does not seem to be a correlation between these line profile variations and the disk-building or dissipation. However, especially in the photospheric \ion{Si}{3} line, we see that there are fast-moving prograde (blue-to-red) and retrograde (red-to-blue) features that occur at the same time in this star. {These features are seen in Fig.~\ref{fig:fast_vary} as absorption components that traverse the line profile on the time-scales of hours.} The width of the subfeatures {is line dependent due to the mass of the atoms making up each line}.

\section{Discussion}
While $\lambda$ Pav is a classical Be star, it seems to spend a large fraction of time without any detectable circumstellar material relative to other early-type Be stars. From an observational standpoint, this property makes it relatively easy to distinguish between stellar versus circumstellar signals, which can be difficult for highly active Be stars with strong disks that are fed with frequent mass ejections. In particular, during intervals of time where there is no detectable emission, all observed signals can be presumed to originate purely from the star. There are several distinct types of variability exhibited by $\lambda$ Pav that were sampled by the data analyzed in this work, including signals originating from the stellar photosphere and the circumstellar environment. These are discussed in the following sub-sections.

\subsection{Low-frequency stellar variability} \label{sec:diff_freq}

Besides the single outburst, the most conspicuous variability is associated with a 6.13 d period ($\nu_1 = 0.163$ d$^{-1}$). This signal appears in TESS photometry as a sinusoidal modulation of brightness with an amplitude of about 15 -- 20 ppt (Figs.~\ref{fig:EW},~\ref{fig:FT}), and is present in all available TESS sectors (spanning about six years between sector 13 and 94). This same signal is also clearly seen in spectroscopic line profile variations, as first noticed by \citet{2011A&A...533A..75L}, and as shown in the dynamical spectra in Fig.~\ref{fig:slow_LPVs}. 

This 0.163 d$^{-1}$ signal roughly corresponds to the difference between two frequencies observed by TESS, at 
1.644 d$^{-1}$ and 1.485 d$^{-1}$ (differing by $\sim$2\%). Such ``difference frequencies'' are routinely observed in Be stars, especially with the recent proliferation of high-precision space photometry \citep[e.g.,][]{2016A&A...588A..56B, Baade2017, 2018pas8.conf...69B}. In general the amplitude of difference frequencies is often found to vary significantly over time. This is not the case for $\lambda$ Pav -- the difference frequency is remarkably regular over the so-far six-year TESS baseline (Fig.~\ref{fig:FT}), and possibly even much longer considering the spectroscopic detection of this signal in data from 1999 to 2001 \citep{2011A&A...533A..75L}, and the photometric detection in Hipparcos data collected in the early 1990s at a frequency of 0.16179 d$^{-1}$ and an amplitude of 0.016 mag \citep[approximately 16 ppt;][]{2002MNRAS.331...45K}. {The high observed amplitude of the low-frequency signal relative to all other periodic signals does not conflict with the interpretation that it is a difference frequency. Combination frequencies (including difference frequencies) can easily have higher photometric amplitudes than their parent signals due to differences in geometry/surface patterns  \citep{2015MNRAS.450.3015K}.}

The line profile variations phased to the period associated with the difference frequency appear to move in a prograde fashion, from blue to red (Figs.~\ref{fig:spec_line_profiles},~\ref{fig:slow_LPVs}). From the stellar parameters given in \citet{2016A&A...595A.132Z}, the rotation frequency of $\lambda$ Pav is 0.46 $\pm$ 0.15 d$^{-1}$, and the critical rotation frequency is 0.62 $\pm$ 0.14 d$^{-1}$ (Paper I). Since the frequency observed at 0.163 d$^{-1}$ is obviously lower than the rotation frequency, it must be retrograde (unless it is zonal, i.e. $m=0$, but this seems inconsistent with the LPV pattern). The pattern that moves across the line profile is relatively simple, appearing similar to an $|m| =$ 1 or 2 mode, as are regularly seen in Be stars \citep{2003A&A...411..229R}, but at a much lower observed frequency. {This low frequency is outside of the range observed in pulsating near-main-sequence B-type stars and the model predictions for them, but as mentioned previously, this is a difference frequency in this star. What is unusual for $\lambda$ Pav is that the difference frequency appears to be extremely stable.} 

Another prominent example of a Be star with a difference frequency is HD 6226. In this case, the difference frequency is not observed directly (either in photometry or line profile variations), but rather is presumed because of two closely spaced pulsation frequencies detected in photometry separated by about 0.0115 d$^{-1}$ (the corresponding period is 87 d). In HD~6226, outbursts occur with an 87 d period (although some events are skipped or this is modulated on two frequencies). This suggests that the timing of the outbursts is dictated by the difference frequency. In $\lambda$ Pav, the difference frequency 
is over an order of magnitude faster than in HD~6226. Further, mass ejections clearly do not occur every $\sim$6 days in $\lambda$ Pav. The situation in $\lambda$ Pav invokes the idea of ``hierarchical clocks'' as discussed in \citet{2018pas8.conf...69B, 2018A&A...610A..70B}, where some unknown mechanism (possibly related to very low difference frequencies) `opens a valve' to allow for mass ejection, where the difference frequency then is able to pump out material {for some period of time. Here, the amplitude, which is much higher than those of single-mode stellar pulsations, could be increased by continuous pumping of matter above the photosphere to allow for the disk building to occur}. In HD~6226, this hypothetical valve is typically open, while in $\lambda$ Pav it is not. One interesting hypothesis that we do not have the ability to test with the current data is that the 0.163 d$^{-1}$ signal increases in its photometric amplitude at the onset of an outburst as may be seen in Fig.~\ref{fig:EW}. However, the TESS data only cover half of one period of this, so it is impossible to gauge, {but similar changes are observed with 25 Ori \citep[B1Vne; ][]{2018pas8.conf...69B}}.

The LPV pattern of the slow 0.163 d$^{-1}$ signal seems qualitatively different from that of the faster 1.644 d$^{-1}$ signal, most clearly seen in Fig. 5 of \citep{2011A&A...533A..75L}, but also in the right panels of Fig.~\ref{fig:slow_LPVs}. For the 0.163 d$^{-1}$ signal, the dark/light patches travel across the line from phase 0 to 1, while for the 1.644 d$^{-1}$ signal the dark patch (the most evident feature in \citealt{2011A&A...533A..75L}) and light patch (the most evident feature in {the middle column of} Fig.~\ref{fig:slow_LPVs} in the pre-outburst data) travel across the line profile over just half of the pulsational phase. 

{It is worth emphasizing one major shortcoming of the TESS photometry and the spectroscopic data analyzed in this work and in \citet{2011A&A...533A..75L}. The frequency resolution, which is dictated by the time baseline, is poor. There may therefore be multiple closely spaced frequencies that we detect as only one signal, especially considering that rapid rotation can lead to very small separations between the frequencies associated with consecutive radial orders \citep[e.g.,][]{2017A&A...598A..74P, 2022MNRAS.511.1529S}. This would also naturally lead to variable amplitudes and/or frequencies in the signals measured by TESS, as unresolved signals beat against each other. }

{That the LPVs and photometric variations from this difference frequency are so clearly observed and are apparently stable for long periods of time makes $\lambda$ Pav an excellent laboratory to study the non-linear mode coupling that seems responsible for the difference frequency \citep{2024A&A...687A.265V}. We are not aware of any other Be star in which the spectroscopic signature of a difference frequency is so evident.
A more detailed and systematic investigation of the line profile and brightness variations associated with the 0.163 d$^{-1}$ signal as well as the two likely `parent' modes (at 1.644 d$^{-1}$ and 1.485 d$^{-1}$) may help to answer several important question: Is the pulsation geometry the same for all three observed signals? Why is the difference frequency so regular compared to the majority of Be stars? How do the observed variations map to features on the stellar surface? Given the early spectral type (B2Ve), which is relatively hot compared to the blue edge of the SPB instability strip, what mechanisms excite the two parent modes?}

\subsection{High-frequency stellar variability} \label{sec:fastLPVs}
Very fast-moving, highly structured LPVs are seen in the dynamical residual spectra from all nine sequences where $\lambda$ Pav was observed for several consecutive hours (Fig.~\ref{fig:fast_vary}). Before attempting to interpret these, it is useful to outline the observational facts. 

\begin{enumerate}

\item These variations occur on the $\sim$1 to a few hours timescales.

\item The character and details of the LPVs differ from one night to the next. {It may be possible to model these as several periodic LPVs, but that will likely require detailed modeling and further observations to determine the timescale for each component.}

\item Features seem to mostly travel across line profiles from blue to red, but also red to blue, or without apparent motion across the line profile (i.e. features appear and disappear at a near constant velocity). 

\item There are no high-frequency signals {($\geq$ 2 d$^{-1}$)} detected in TESS. 

\item The {LPVs} are in general highly structured. 

\item Fast LPVs were seen during the disk-less phase, during active mass ejection, and during disk dissipation. 

\item The acceleration of features across the line profiles can appear to vary. 

\end{enumerate}

Interpreting the physical cause of these variations is not entirely straightforward, as there are several distinct processes that can cause line profile variations. A star with an inhomogeneous surface (e.g. with temperature and/or chemical spots) can exhibit LPVs with a frequency equal to the stellar rotation (or over a range of frequencies if the star has surface differential rotation) \citep{2006PhDT.......425S}. Such spots may vary over time (appearing, migrating, and disappearing as with Sun spots), or they may be apparently constant over long time baselines. Variations in the circumstellar environment can cause LPVs, too, and these variations can be intrinsic or related to projection effects from obscuration of (parts of) the observed stellar disk and/or self-obscuration as material rotates or orbits around the star. Pulsation is another mechanism, as there are several types of pulsation with many different geometric possibilities which can induce LPVs with a wide range of patterns and timescales \citep{2003Ap&SS.284...85T}. 

It is useful to keep in mind that a typical set of continuous observations in our data is $\sim$5 hours long, which is approximately $\sim$10\% of the rotation period, or $\sim$20\% of the near-surface orbital period \citep[both of these estimated from the parameters of][]{2016A&A...595A.132Z}.

Of the nine nights of continuous observations, three were about two weeks before the outburst in consecutive nights (``set A''), two were acquired a few days before the outburst (with a gap of one night between them; ``set B''), two were acquired on consecutive nights during the build-up phase of the outburst (``set C''), and two were obtained on consecutive nights about 10 days after the outburst peaked in EW (``set D''). 

In \ion{Si}{3} $\lambda$5739, which is the only line apparently unaffected by circumstellar material, there do not seem to be any major differences in the appearance of the fast-moving features between sets A -- D. All variations are confined to within $\pm200$ km s$^{-1}$, and a combination of prograde and retrograde features {is} seen. These LPVs bear resemblance to high-order (in terms of $|m|$) pulsation \citep{1994PhDT.......131R}. 

It is not uncommon for early-type Be stars to exhibit high-frequency p-modes {\citep[i.e. $\beta$ Cephei pulsation with frequencies between $\sim 3$ and $\sim 14$ cycles d$^{-1}$][]{2005ApJS..158..193S}}. The characteristics of the fast LPVs in $\lambda$ Pav seem different for the following reasons. For example, the early-type Be stars $\pi$ Aqr and BZ Cru both exhibit high-frequency p modes, which i) are detected in photometry, and ii) appear as regular periodic patterns in line profiles \citep{2020MNRAS.494..958N}. In $\lambda$ Pav, there are no corresponding photometric frequencies {in Fig.~\ref{fig:FT}} (despite the high precision of TESS, although geometric cancellation can make high-order $|m|$ modes all but invisible to photometry), and the LPV patterns differ significantly from night to night, {as expected for multi-mode pulsations}. For instance, on the night of 2023 July 28 {(row 7 in Fig.~\ref{fig:fast_vary})}, four or five features move roughly in parallel from blue to red, but possibly with changing acceleration throughout the night. On 2023 Aug 12 {(row 8 of Fig.~\ref{fig:fast_vary})}, there are only two or three features, primarily only on the red half of the line, which hardly move across the line profile but rather appear and disappear at near-constant velocity. This could represent a crossing of prograde and retrograde moving features at that time. On 2023 July 22, {(row 4 of Fig.~\ref{fig:fast_vary})}, there are again four or five features, but moving with clearly different accelerations, most being prograde, but some being retrograde. The work of \citet{2016A&A...595A.132Z} finds that $\lambda$ Pavonis has an inclination of 36$^\circ$ $\pm$ 10$^\circ$ and a $v \sin i$ of 145 km s$^{-1}$ $\pm$ 12 km s$^{-1}$. 
Considering the low inclination angle, the retrograde moving features could actually be propagating in the direction of stellar rotation, but are visible from the ``back'' side of the star and thus move from red to blue in the observed line profiles.

The TESS observations show no indication of photometric signals associated with these fast LPVs, indicative that these should be higher-order LPVs. As a first attempt, we modeled several different combinations of pulsation geometries, and found the closest qualitative match was with an $l$ = 7, $m$ = -7  sectoral prograde mode in BRUCE/KYLIE \citep{2014ascl.soft12005T} as the observed frequency is coursely proportional to $|m|$. The resulting variations qualitatively match the dominant features in \ion{Si}{3} $\lambda$5739 on the night of 2023 July 22 (Fig.~\ref{fig:fast_vary}{, column 4, row 4}). However, this single mode of course cannot explain the night-to-night differences or the sometimes retrograde or near-stationary features. Like with the slow LPVs, a dedicated effort to model these features is needed to understand their details and is beyond the scope of this paper. It is possible that multiple coherent modes are present at the same time, constructively and destructively interfering in ways that at first glance appear random, but could be understood if the constituent mode frequencies and geometries are known. Alternatively, the implied high-order pulsation could be non-coherent. It is noteworthy that these fast-moving relatively sharp features do not seem to appear in any of the spectra presented in \citet{2011A&A...533A..75L}. We do not doubt the reality of these signals in the 2023 observing run, since they are clearly seen {with both NRES and CHIRON} in every long observing sequence. 

Qualitatively similar features have been observed in a few other early-type Be stars, including $\gamma$ Cas and $\zeta$ Tau \citep{1988PASP..100..233Y, 1990ApJS...74..595Y, 1999ApJ...517..866S, 1998ApJ...507..945S}. 
In both of these cases, the stars possessed strong disks at the time of observations, so that both stellar and circumstellar interpretations {could be} valid. However, for $\lambda$ Pav, it is safe to attribute these to pulsational features on the stellar surface thanks to observations of the purely photospheric \ion{Si}{3} $\lambda$5739 line and the absence of any circumstellar material prior to the outburst. 

During the rising phase of the outburst (2023 July 27, 28, {rows 6 and 7} in Fig.~\ref{fig:fast_vary}), the \ion{H}{1} and \ion{He}{1} lines, which are sensitive to circumstellar material, show clear variations well outside of the stellar $v$sin$i$, especially evident on 2023 July 27 on the red side of the line. As discussed in Sec.~\ref{sec:obs_outburst}, the circumstellar material is still inhomogeneously distributed at these early times prior to circularization. While the underlying pulsational variability makes it difficult to track the motion of any gas during these several hour sequences, the super-$v \sin i$ features (which are not observed in the photospheric \ion{Si}{3} $\lambda$5739 line) can safely be attributed to circumstellar material. While this would nominally violate the conservation of angular momentum, in these cases, the material is conserving angular momentum by orbiting the star {by being viscously advected to larger distances}. Recent simulations of inhomogeneous mass ejection leading to disk build-up in Be stars show that prior to circularization, and while mass ejection is ongoing, the gas dynamics are complex and do not resemble circular Keplerian orbits (Rubio et al., {priv. comm.}). High-cadence spectroscopic observations during the early stages of mass ejection such as for $\lambda$ Pav will be important in refining models of this sort to better understand the initial conditions of ejected material, hopefully elucidating the stellar processes that give rise to the ejection of mass. 

\subsection{Circumstellar environment}

During our intensive spectroscopic observing campaign, and overlapping with space photometry from TESS in June and July 2023, the $\lambda$ Pav system transitioned from a disk-less state by means of a discrete mass ejection episode {to} the formation of a new disk. The disk build-up phase lasted for only around 3--5 days for all lines sensitive to emission (Fig.~\ref{fig:EW} and the exponential fit in Table~\ref{tab:fit_params}), during which time the emission levels of \ion{H}{1} and \ion{He}{1} lines increased while emission asymmetry oscillated with a period of about 1.25 d ($\nu=0.8$ d$^{-1}$; Fig.~\ref{fig:VRfit}, Table~\ref{tab:fit_params}). After peak emission levels were reached, the dissipation proceeded more quickly for \ion{He}{1} lines compared to \ion{H}{1}. About 30--40 days after the start of the event, emission features were no longer present in any lines at optical wavelengths. Only the first few days of the event were captured by TESS, so the timescale of the build-up and dissipation phases in continuum flux remain unknown. Overall, the behavior related to the short outburst in $\lambda$ Pav is much the same as in comparable events in other Be stars (Paper I).

The observed behavior of the emission levels of the \ion{H}{1} and \ion{He}{1} lines is consistent with the generally accepted picture of Be disk build-up and dissipation. During the build-up phase, the inner volume of the disk fills up relatively quickly. At these early times, all observables originate from near to the star (insufficient time has elapsed for material to be viscously transported to large distances), and we observe the same build-up times in both the \ion{H}{1} and \ion{He}{1} emission. When mass ejection turns off, the inner-most regions of the disk dissipate more quickly with the bulk of the recently ejected material falling back on to the star \citep[e.g.,][]{2021ApJ...912...76M, 2021ApJ...909..149G}. Helium emission tends to arise from hot dense material relatively close to the star, so He emission levels drop relatively quickly during the dissipation phase. Furthermore, we see that the higher-{excitation} \ion{He}{1} at 4713 and 4921 \AA\ dissipate quicker than the lower-{excitation} \ion{He}{1} 5876 and 6678 lines. Hydrogen emission, on the other hand, can arise from cooler and less dense regions of the disk farther from the star. All relevant {time dependent events}, in particular the rate of change in density caused by variations in the stellar mass ejection rate, proceed more slowly at larger radial distances \citep{2012ApJ...756..156H}. Thus, the slower drop in \ion{H}{1} emission, relative to \ion{He}{1}, is consistent with the picture of Be disks dissipating `inside-out'. In this picture, the density of the inner disk drops quickly, while the decreasing density in more outer parts of the disk lags behind. In fact, this behavior can be directly observed for H$\alpha$ in Fig.~\ref{fig:spec_line_profiles}. The high-velocity wings of H$\alpha$, which originate in the more inner disk region, dissipate more quickly than the emission closer to the line center which preferentially forms at larger radii in the disk where Keplerian orbital speeds are lower.

\section{Conclusions}

The findings in this paper have detailed pulsational properties both photometrically and spectroscopically at times when a disk was built and dissipated around the classical Be star $\lambda$ Pav. The data set presented will provide theorists a rich landscape with which to model how small ejections enter the Be star's disk and then dissipate with time. Furthermore, the pulsational properties of the star will be important to understand this star's properties and how they relate to the disk evolution. Lastly, some prevailing thoughts for Be stars may need to be reconsidered given the way in which the fast line profile variations are {not} similar at the times before, during, and after the outburst. {Our main findings are as follows:

\begin{enumerate}

\item At the end of two sectors of monitoring with TESS, the bright Be star $\lambda$ Pav began to grow a disk from a pristine diskless state. The evidence of the disk formation was an increase in TESS flux as well as emission seen in Balmer lines and optical \ion{He}{1} lines.

\item The timescale for disk build-up was consistent in all lines that were analyzed ($\sim$4 days; Sec.~\ref{sec:obs_outburst}). After reaching peak emission levels, the decay proceeded more quickly for \ion{He}{1} lines ($\sim$4 days) compared to H$\beta$ ($\sim$6 days) and H$\alpha$ ($\sim$9 days). 

\item During the first several days of the outburst, the emission asymmetry oscillated with a  period of $\sim$1.25 days in all measured lines with emission features (Sec.~\ref{sec:obs_outburst}), {which we interpret as orbiting circumstellar material}. After $\sim$10 days from the start of the event, emission features had settled into a symmetric profile. 

\item Besides the outburst, a signal with a 6.13 day period is the most prominent variability in the TESS photometry. A spectroscopic signal at this period was noted in \citet{2011A&A...533A..75L}, and is also detected in the LPVs of our spectra. The frequency of this signal is consistent with the difference between two higher-frequency signals (at 1.644 d$^{-1}$ and 1.485 d$^{-1}$), {which could be NRPs}. The variations associated with this difference frequency are remarkably stable, apparently over many years (Sec.~\ref{sec:diff_freq}), and also, {spectroscopically,} before, during, and after the outburst.

\item Fast-moving, highly structured, and apparently changing features were evident in line profiles in all nine nights where $\lambda$ Pav was observed for several consecutive hours (Sec.~\ref{sec:fastLPVs}). These appear as a combination of prograde and retrograde features, were seen before, during, and after the outburst, and have no photometric counterpart.

\item \citet{2011A&A...533A..75L} found four frequencies in their spectroscopic analysis of $\lambda$ Pav. We confirm that three of these {(0.163 d$^{-1}$, 0.82 d$^{-1}$, and 1.644 d$^{-1}$)} are present in the TESS photometry. 

\end{enumerate}

\begin{acknowledgments}

We thank our referee, Dietrich Baade, who provided comments that improved this paper. We also thank Amanda Rubio for comments that led to some of the discussion in the paper.
S. S. N. and N. D. R. are grateful for support from NASA grants 80NSSC23K1049 and 80NSSC24K0229.
S. S. N. and S. G. F. are grateful for support from Embry-Riddle Aeronautical University's Undergraduate Research Institute and the Arizona Space Grant Consortium.

This work makes use of observations from the LCOGT network. This paper includes data collected by the TESS mission, which are publicly available from the Mikulski Archive for Space Telescopes (MAST). Funding for the TESS mission is provided by NASA’s Science Mission directorate. This research has used data from the CTIO/SMARTS 1.5m telescope, which is operated as part of the SMARTS Consortium by RECONS (www.recons.org) members T. Henry, H. James, W.-C. Jao, L. Paredes, S. Carrazco-Gaxiola, and T. Johns. 
J.L.-B. acknowledges support from the European Union (ERC, MAGNIFY, Project 101126182). While partially funded by the European Union, views and opinions expressed are however those of the author only and do not necessarily reflect those of the European Union or the European Research Council. Neither the European Union nor the granting authority can be held responsible for them.

\end{acknowledgments}

\begin{contribution}

All authors contributed to this writing. SSN led the analysis and measurements and prepared most of the manuscript under the guidance and mentorship of NDR. JLB provided guidance in the analysis and data, also scheduling the LCO observations and leading the analysis of the TESS light curve. NDR scheduled and planned the CHIRON observations. SGF provided the foundation of the MCMC code used in this manuscript.  

\end{contribution}

\facilities{LCOGT (NRES), CTIO:1.5m (CHIRON), TESS}

\software{astropy \citep{2013A&A...558A..33A,2018AJ....156..123A,2022ApJ...935..167A}
          }

\bibliography{sample631}{}
\bibliographystyle{aasjournalv7}

\appendix

\begin{figure}
    \centering
    \includegraphics[width=0.95\linewidth]{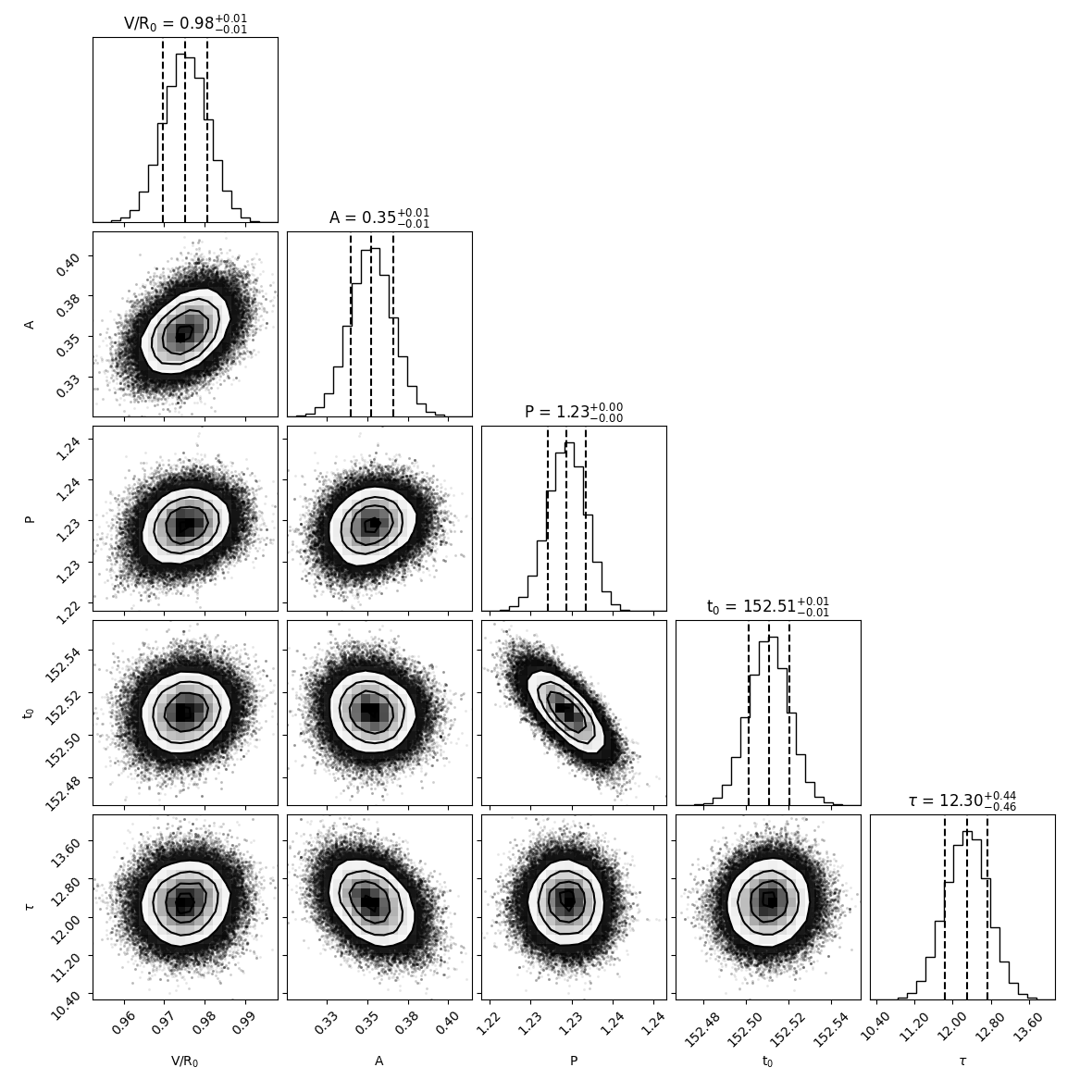}
    \caption{An example corner plot showing the fitting for the V/R circularization for the H$\alpha$ line.}
    \label{fig:corner}
\end{figure}

\end{document}